\documentclass[pre,aps,showpacs,twocolumn,floatfix,superscriptaddress]{revtex4}
\usepackage[english]{babel}
\usepackage{inputenc}
\usepackage{amsmath}
\usepackage{cancel}
\usepackage{array}
\usepackage{psfrag}
\usepackage{graphicx}
\usepackage{bm}

\begin{document}
\title{Modeling the mechano-chemistry of the $\phi 29$ DNA translocation motor}

\author{R. Perez-Carrasco}
\affiliation{Departament d'Estructura i Constituents de la Mat\`eria, Facultat de F\'isica, Universitat de Barcelona C/Mart\'i Franqu\'es 1, 08028 Barcelona, Spain.}

\author{A. Fiasconaro}
\affiliation{Dpto. de F\'{\i}sica de la Materia Condensada,
Universidad de Zaragoza. 50009 Zaragoza, Spain}
\affiliation{Instituto de Ciencia de Materiales de Arag\'on,
C.S.I.C.-Universidad de Zaragoza. 50009 Zaragoza, Spain.}

\author{F. Falo}
\affiliation{Dpto. de F\'{\i}sica de la Materia Condensada,
Universidad de Zaragoza. 50009 Zaragoza, Spain}
\affiliation{Instituto de Biocomputaci\'on y F\'{\i}sica de Sistemas
Complejos, Universidad de Zaragoza. 50009 Zaragoza, Spain}

\author{J. M. Sancho}
\affiliation{Departament d'Estructura i Constituents de la Mat\`eria, Facultat de F\'isica, Universitat de Barcelona C/Mart\'i Franqu\'es 1, 08028 Barcelona, Spain.}

\begin{abstract}
We present a study of  the DNA translocation of the bacteriophage $\phi 29$ packaging molecular motor. From the experimental available information we present a model system based in an stochastic flashing potential, which reproduces the experimental observations such as: detailed trajectories, steps and substeps, spatial correlation, and velocity. Moreover the model allows the evaluation of power and efficiency of this motor. We have found that the maximum power regime does not correspond with that of the maximum efficiency. These informations can stimulate further experiments.

\end{abstract}


\maketitle

\section{Introduction}

Molecular motors, discovered in living cells almost 30 years ago,  constitute essential ingredients for life, and  have been extensively studied since then with two approaches: biochemical and/or physical~\cite{Schliwa, mickler,fisher}.
Many motor are ATP fueled like kinesins and dyneins, which can ``walk" along a cellular microtubule and perform transport features inside cells. Other examples include myosins, able to contract the muscles tissues with a strongly synchronized movement; or translocator motors, which move biopolymers across membranes~\cite{RMP,Wendell}.
Other motors are driven by an ion flux such the bacterial flagella motor (BFM), which propels the bacteria in a fluid media.

Many important cellular processes use the physical mechanism of the translocation of biomolecules through cell membranes~\cite{RMP,Wendell} such as proteins, RNA and DNA strands.
The interest in this subject is twofold: understanding the relevance of this mechanism of living systems, but also to investigate the enormous technological possibilities of controlling it by using synthetic materials able to imitate this important biological function ~\cite{mickler}.
In this sense, the passage of biomolecules through nanopores involves many disciplines, from biology to nanothechnology, such as the passage of mRNA through nuclear pores \cite{KasPNAS96} or the translocation of DNA in graphene pores
\cite{Gregory}.

In many cases, translocation is driven by constant and/or noisy forces inside the pores or by the chemical potential difference between both sides of the membrane. In other cases, the process is assisted by an intermembrane ATP-based molecular motor \cite{Mehta}.

To this last category belongs the motor of the $\phi 29$ bacteriophage which performs the translocation of its DNA inside the virus capsid (packaging). This motor translocates DNA molecules against the huge pressure inside the virus capsid, which can achieve 40 atm at the end of the process, making it one of the strongest molecular machines known. The experimental study of this motor, conducted mainly by the Bustamante group \cite{Bust01,Bust05,Bust09}, allowed the knowledge of the DNA dynamics during the translocation at a very detailed either time or spatial scales.

Recently a model has been presented in order to describe the main features of the $\phi 29$ bacteriophage translocation dynamics \cite{ajf-damn}, as a biological extension of a time varying driven translocation of long molecules \cite{ajf-rtn,ajf-sin}.
The idea of the model is to drive with a constant force a polymer chain in one direction, while in its activated state, but it leaves the polymer diffuse freely when the motor is inactive
\cite{starikov,sancho,ajf-damn}.
The mechanism is the origin of the Michaelis-Menten (MM) polymer velocity, related to a microscopic re-interpretation of the MM enzymatic reaction~\cite{ajf-damn}.

Although the detailed mechanism of the inner structure of the motor remains to be better clarified, a sub--stepping feature has been also revealed experimentally~\cite{Bust09}. These experiments show that each cycle of the motor consists of two well differentiated processes. First a purely catalytic process in which around four ATPs are bound to the pentameric ring. During this phase, the DNA does not advance. Hence, this process is referred as the dwell phase. The second process of the cycle is the burst phase in which the four ATPs are hydrolyzed with a total DNA advance of $\sim$10 bp (3.4 nm). Slowing down the hydrolysis be means of an external hindering force, a finer discretization of the main step is found obtaining four substeps of $\sim$2.5 bp (0.85 nm) each one, coinciding with the number of ATP hydrolysis processes.

The present study points its attention to the specific sub--step structure of the trajectories, here modeled as sub--activation states of the simpler ATP machine depicted in \cite{ajf-damn,sancho}. As it has been already justified in there, for the experimental constraints and the small dimension of the motor, the translocation can be approximately described by a rigid chain pushed in a 1--dimensional domain.

In the next section all the details of the model are described and correlated with the biochemical available information. In Section III we present the numerical results with the corresponding analysis and comparison with the experiments. Finally we end with some comments and conclusions.

\section{The Model}

The full experimental set up consists of the packaging motor, the DNA strand and two polystyrene beads of 860 nm of diameter. However, since the strand is always kept stretched, its elastic motion can be neglected. Thus, the problem can be considered as a solid motor--bead system advancing along a DNA track. The dynamics of such a nanoscopic system can be described by an overdamped Langevin equation,

\begin{equation}
\gamma \dot x = -V'(x,t) + F_E + \xi(t),
\label{Langevin}
\end{equation}
being $\gamma$ the friction coefficient carried in the motion of the motor, $x$ the relative position of the motor along the track, $V(x,t)$ the motor potential describing the working process of the motor, $F_E$ the external force applied on the motor and $\xi(t)$ the thermal force which can be described as white Gaussian noise of zero mean and correlation,

\begin{equation}
\langle\xi(t)\xi(t') \rangle=2\gamma kT \delta(t-t').
\end{equation}

The value of $\gamma$ used in the current analysis (Table \ref{table}) is an heuristic extrapolation from the observed experimental data.  All the biochemical relevant  information of the working cycle of the motor, as it is described in Refs. ~\cite{Bust09} is allocated in the flashing potential $V(x,t)$. Since the working of the motor presents two different regimes it is expected that two different potentials apply $V_D(x)$ and $V_B(x)$ for the dwells and the burst phases respectively. Therefore, the temporal and the spatial dependence of the potential can be separated as,

\begin{equation}
V(x,t)=V_D(x) + \left( V_B(x)-V_D(x) \right) \eta(t),
\end{equation}
where $\eta(t)={0,1}$ sets the state of the motor potential. $\eta_D=0$ for the dwell phase and $\eta_B=1$ for the burst phase. This description introduces a flashing mechanism for the state transition and implies that the change in conformation occurs in a faster timescale than the actual motion of the strand. This difference of timescales has been reported for several motors~\cite{sancho,sancho2}.

\begin{figure}
\includegraphics[width=1\columnwidth]{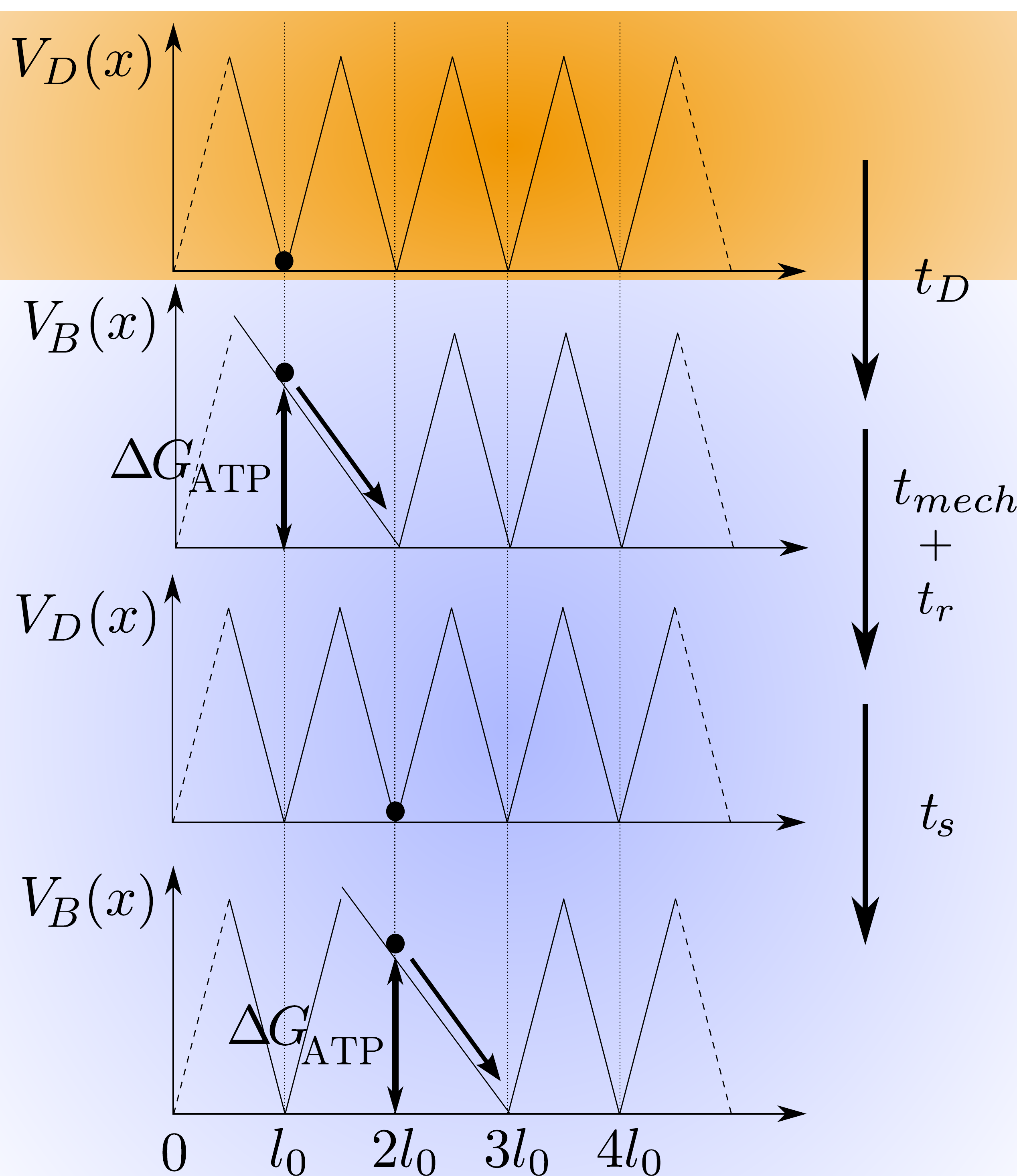}
\caption{\label{potential} (Color Online) Flashing potential describing the motor cycle. The cycle starts with the dwell phase (orange). Once the four ATP are bound, the burst phase (blue) begins. The cycle  is not fully depicted, in order to complete the cycles two additional substeps are necessary.}
\end{figure}

The structure of each potential landscape can be reconstructed from experimental results ~\cite{Bust09}. For the dwell state, there must be a minimum for each equilibrium position that keeps the DNA fixed when no ATP is present in the system and an external force pulls the strand. A possible scenario is that of a periodic potential with the periodicity of a substep $l_0\!=$2.5 bp. The simplest potential with such a periodicity is a ratchet potential (Fig. \ref{potential}). Its height, $V_D^\mathrm{MAX}$, sets the magnitude of the fluctuations around the dwell state. However, there are two limitations to its value. On one hand the height of the potential must introduce a force that is larger or comparable to the stall force of the motor $\sim 80$ pN. Conversely, a very energetic potential may artificially introduce extra energy in the system through the potential flashing. The values used in the current work are gathered in Table (\ref{table}).

On the other hand, the excited or burst potential must introduce a net force which drives the motor  to the next minimum of the relaxed potential. This force results from the new form of the potential and it can be related with a change in the structure of the protein that modifies the equilibrium positions of the strand. Since the potential is determined by the motor structure, this modification is confined to a region of space, \textit{i.e.} only a region of the $V_D(x)$ is modified when switching to $V_B(x)$ (see Figure \ref{potential}). Note that this scenario is different to the introduction of a constant force with no spatial limitations where the motion and the energy supplied are uncontrolled.

The height of the burst potential $V_0$ is not a free parameter but is determined by the free energy used from the hydrolysis of a molecule of ATP (Fig \ref{potential}). Obtaining,
\begin{equation}
\frac{V_B^\mathrm{MAX}}{\frac{3}{2}l_0}=\frac{\Delta G_{\mathrm{ATP}}}{l_0}\quad\Rightarrow\quad V_B^\mathrm{MAX}=\frac{3}{2}\Delta G_{\mathrm{ATP}}.
\end{equation}

Finally, it is necessary to set the temporal dynamics of the motor. The easiest description of the dwell time $t_D$ is the necessary time to take irreversibly four ATPs independently with a constant reaction rate $k_\mathrm{ATP}$ \emph{i.e.} the stochastic time resulting of adding four exponential events. Hence, the resulting probability density is the Erlang distribution,

\begin{equation}
P(t_{D})=\frac{k_\mathrm{ATP}^4t_{D}^3\mathrm{e}^{-k_\mathrm{ATP}t_D}}{3!}.
\end{equation}

Which has an average time $\langle t_D \rangle = \frac{4}{k_\mathrm{ATP}}$ and returns a randomness
\begin{equation}
\rho=\frac{\langle t_D^2\rangle}{\Delta t_D^2}=4.
\end{equation}

Experimentally, an hyperbolic dependence of the reaction rate $k_{\mathrm{ATP}}$ has been observed an characterized with rates $k_{\mathrm{ATP}}^0$ and $t_{\mathrm{ATP}}^0$. This description is in agreement with the classic Michaelis-Menten theory and it allows to write  $k_{\mathrm{ATP}}$ as,

\begin{equation}
k_{\mathrm{ATP}}=\left(\frac{k_{\mathrm{ATP}}^0}{[\mathrm{ATP}]}+t^0_{\mathrm{ATP}}\right)^{-1}
\end{equation}

On the other hand the burst phase can be split in 4 substeps. Each substep consists of a mechanical process (assisting the advance of the strand) plus a catalytic process (that generate the substep dwells). The stroke duration of each substep can be computed directly from the dissipative dynamics leading the motion,

\begin{equation}
\gamma v_\mathrm{mech}=\frac{\Delta G_{\mathrm{ATP}}}{l_0}\quad\Rightarrow\quad t_\mathrm{mech}=\frac{\gamma l_0^2}{\Delta G}.
\end{equation}

Additionally, another factor remains unknown from the trajectory, which is how the substep dwell is extended along the two conformational states of the protein \emph{i.e.} if the motor stays in the excited conformation more time than $t_{mech}$ prior to the flash to the relaxed state. This time is related to the existence of a process triggering the flashing of the potential that has a scale around the $ms$. Such a factor is very important since it will allow the motor an additional time to reach the minimum increasing this way the coupling ratio of the motor. This  reaction time $t_r$ can be introduced as an extension of the mechanical time $t_{mech}$. For the sake of simplicity, $t_r$ will be introduced deterministically and the effect of its length in the motor performance will be studied in the following section.

On the contrary, the catalytic event after each substep has been experimentally characterized as a leading reaction of a rate $k_{s}(F_E)$ that depends on the external pulling force applied, and presents a probability density
\begin{equation}
P(t_s)=k_s\mathrm{e}^{-k_st_s}.
\end{equation}
The rate $k_s$ at large external pulling forces (-40 pN)  is measured to be $k_{s}=22 \, \mathrm{s}^{-1}$,  while it is very quick at low forces and cannot be measured ($k_s>1\,\mathrm{ms^{-1}}$)~\cite{Bust05}. A possible description for this rate is an exponential rate dependence,
\begin{equation}
k_s=\frac{\mathrm{e}^{-F_E/F_0}}{t^0_s} \label{ks},
\end{equation}
with $F_E \sim - 52 {\mathrm{pN}}$.

The exact dependence of the rate with the external force is not important for the current description as long as $k_s$ keeps the values mentioned above. With the dependence in (\ref{ks})  for low forces, the substep dwell time is $k_s^{-1}(F_E=0)=t^0_s$. Therefore, $t^0_s$ measures the substep dwell time which must be a fast time bellow the millisecond. Once $t_s^0$ is fixed, the rate $k^0_s$ that accomplishes that $k_s(-40 \,\mathrm{pN nm})=22\,\mathrm{s}^{-1}$ can be calculated.

This closes the drawing of the dynamics of a cycle (Fig. \ref{potential}), which it is worth to summarize here. It starts in the dwell state $\eta(t)=0$ and switches to the burst state after a stochastic time $t_D$. After that, the potential starts the first stroke of the cycle flashing to the burst state $\eta(t)=1$ that lasts a deterministic time $t_{mech}+t_r$ flashing back to the dwell potential for a stochastic time $t_{s}$. This process is repeated three more times finishing the burst phase and therefore the full cycle of the motor.

The value of the parameters used has been summarized in Table I.

\section{Results}

\begin{figure*}
\includegraphics[width=1.7\columnwidth]{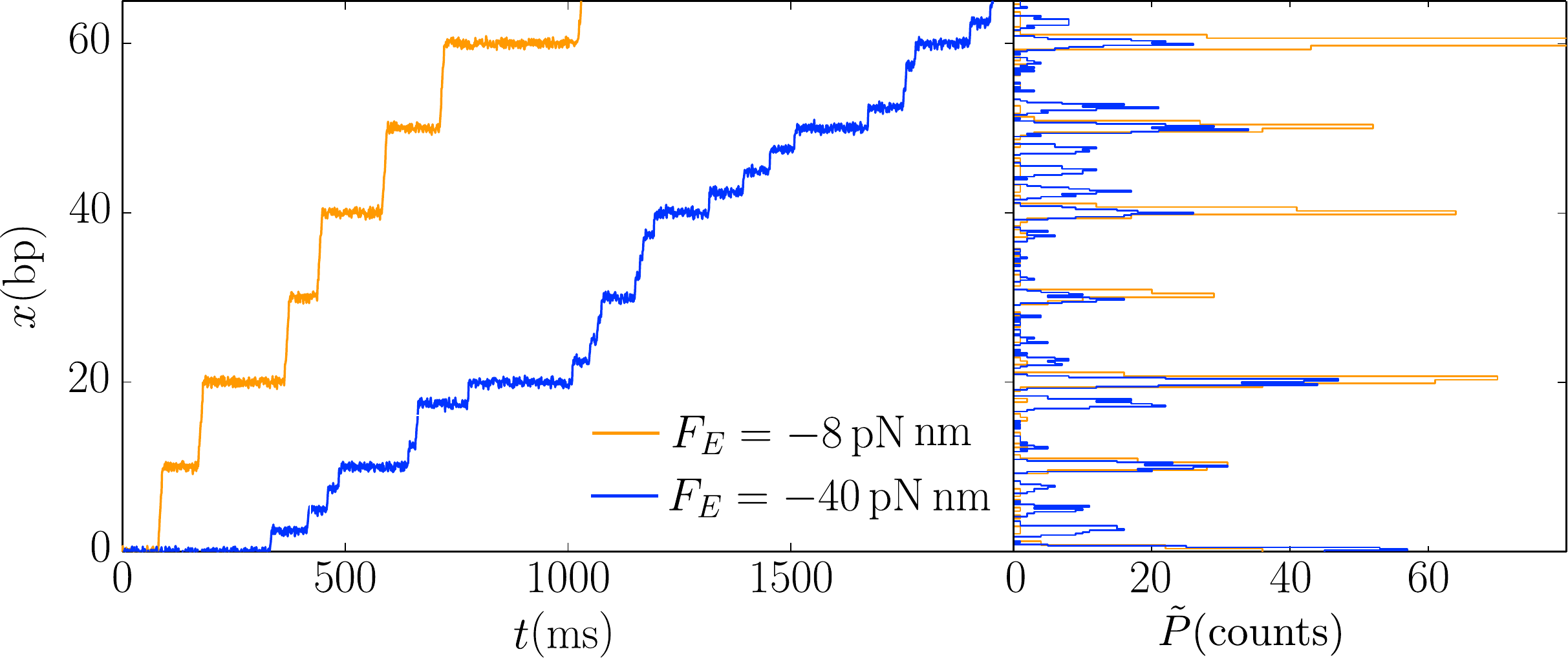}
\caption{\label{traj} (Color Online) Left: Trajectory of the DNA as a function of time for different hindering forces. Right: Corresponding position histogram }
\end{figure*}

Introducing the dynamics and the potential landscape in eq. (\ref{Langevin}) and using the parameters of (Table \ref{table}) the description is complete and the stochastic equation of motion (\ref{Langevin}) can be simulated.

\begin{table}
\begin{tabular}{c|c}
Parameter&Value\\\hline\hline
$l_0$&0.85 nm\\
$\gamma$& 200 pN ms/nm\\
$kT$& 4.1 pN nm\\
$\Delta G_{\mathrm{ATP}}$& 90 pN nm\\
$V_D^\mathrm{MAX}$&150 pN nm\\\hline
$k^0_\mathrm{ATP}$&$650 \, \mathrm{ms\,\mu M}$\\
$t^0_\mathrm{ATP}$&$28.8$ $\mathrm{ms}$\\\hline
$F_s$&$6 \,\mathrm{pN}$\\
$t_{s}^0$& 0.05 $\mathrm{ms}$\\\hline
$t_{r}$& 1 $\mathrm{ms}$

\end{tabular}
\caption{\label{table} Parameters used in the simulation.}
\end{table}

\begin{figure}
\includegraphics[width=\columnwidth]{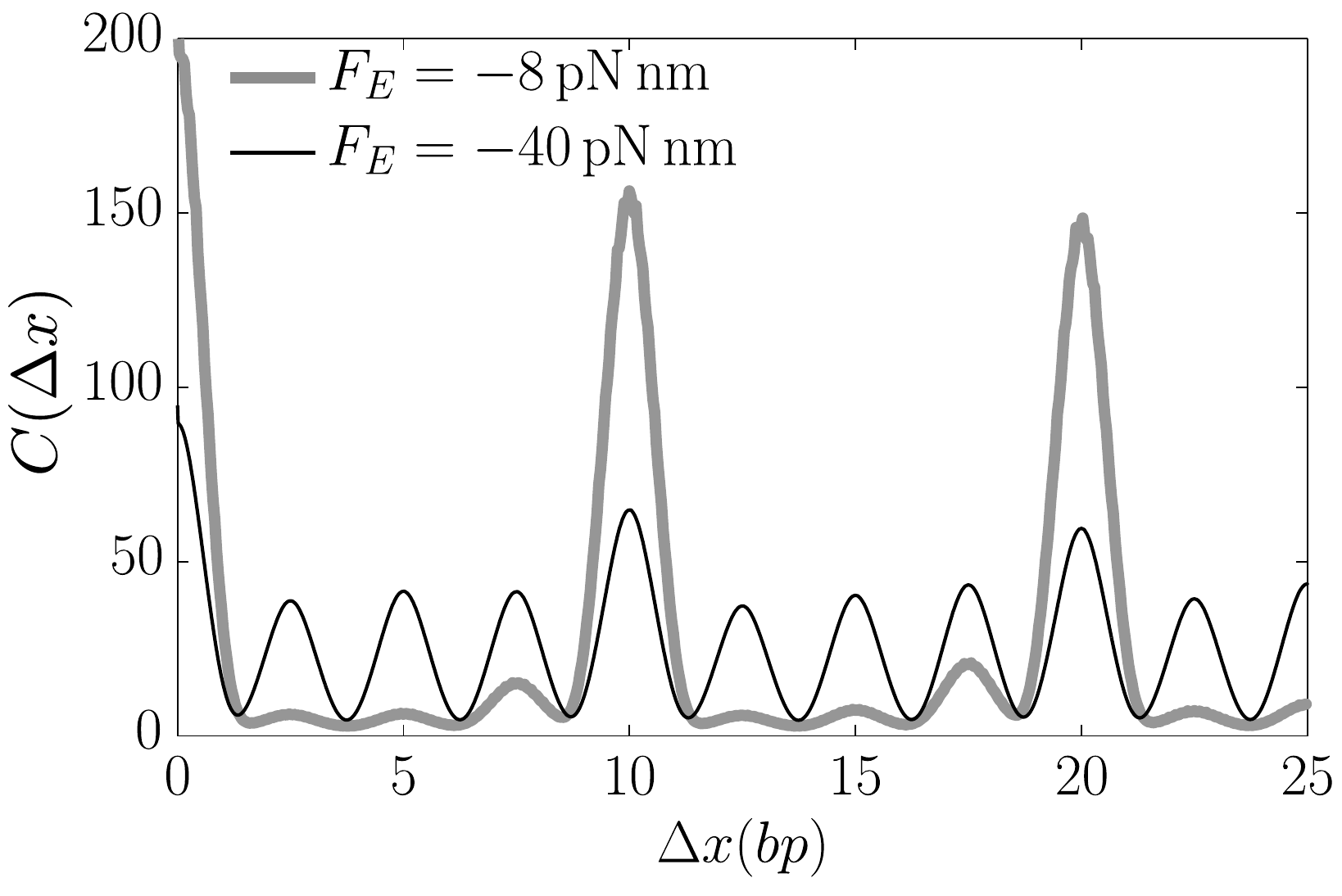}
\caption{\label{cor} Spatial correlation of the two trajectories in Fig. \ref{traj} for a run of 10 seconds.  $F_E=-8\mathrm{pN}$ (top) and $-40 \mathrm{pN}$ (bottom).}
\end{figure}

The resulting trajectories are steplike with a sub--stepping dynamics dependent of the external force. In Fig.~\ref{traj} it is possible to notice that when the pull force is low ($F_E=-8\mathrm{pN}$), then the substep feature of the trajectory is hardly visible, while conversely it is well visible for pulling force of $F_E = -40\mathrm{pN}$. The statistical properties of the substepping can be better studied through a correlation analysis of the trajectories (Fig.~\ref{cor}), obtained as the autocorrelation of the spatial density profile of the motor $\tilde P(x)$,
\begin{equation}
C(\Delta x)=\langle \tilde P(x+\Delta x)\tilde P(x)\rangle_x,
\end{equation}
where $\tilde P(x)$ is computed as the position histogram for the trajectories. The correlation reveals the spatial stepping behavior described. Again, the substepping features are very clear for the strongest pull force, and quite absent for the lowest value. The evidence of the stepping mechanism for higher pull forces coincides with the experimental observations. The two trajectories in Fig.~\ref{traj} can be compared with the  experimental results in Ref.\cite{mof},  Figs. 1b and 3a for $F_E = -8\mathrm{pN}$ and $F_E = -40\mathrm{pN}$ respectively. Analogously, the two lines of Fig.~\ref{cor} represent quite finely the plots in Fig.~1c and 3b of the same article.

\begin{figure}
\includegraphics[width=\columnwidth]{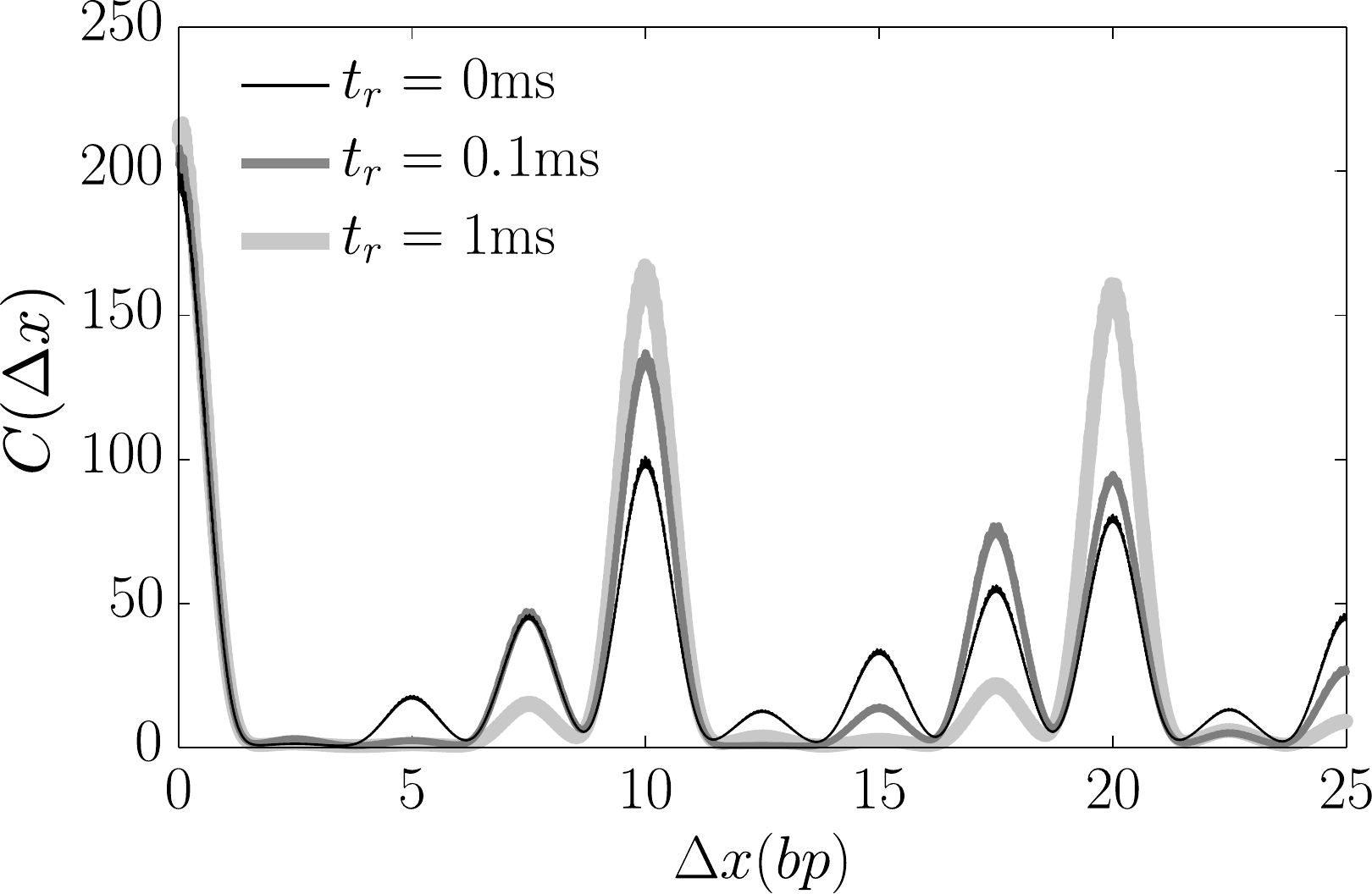}
\caption{\label{figt0} Spatial correlations for three values of $t_r$, namely  $t_r = 0, 0.1$, and $1\mathrm{ms}$. The pull force applied is $F_E = -8\mathrm{pN}$. }
\end{figure}

The trajectories also show that the greater the external opposing force is, the slower is the motor. There are two reasons for this effect. On one hand a greater hindering force reduces the net stroke force of the motor. On the other hand it tilts the potential increasing the number of failed steps. These are the steps where the full ATP hydrolysis energy is consumed but the motor does not arrive at the following minimum of $V_D(x)$. This effect can also be studied through the deterministic time $t_r$ observing that an increase in the reaction time prior to the flashing to the relaxed state brings an increasing of the coupling ratio (Fig. \ref{figt0}). It is interesting to note that the time $t_r$ in all the cases studied here is very small and practically does not increase in any case the time of the cycle.

\begin{figure}
\includegraphics[width=\columnwidth]{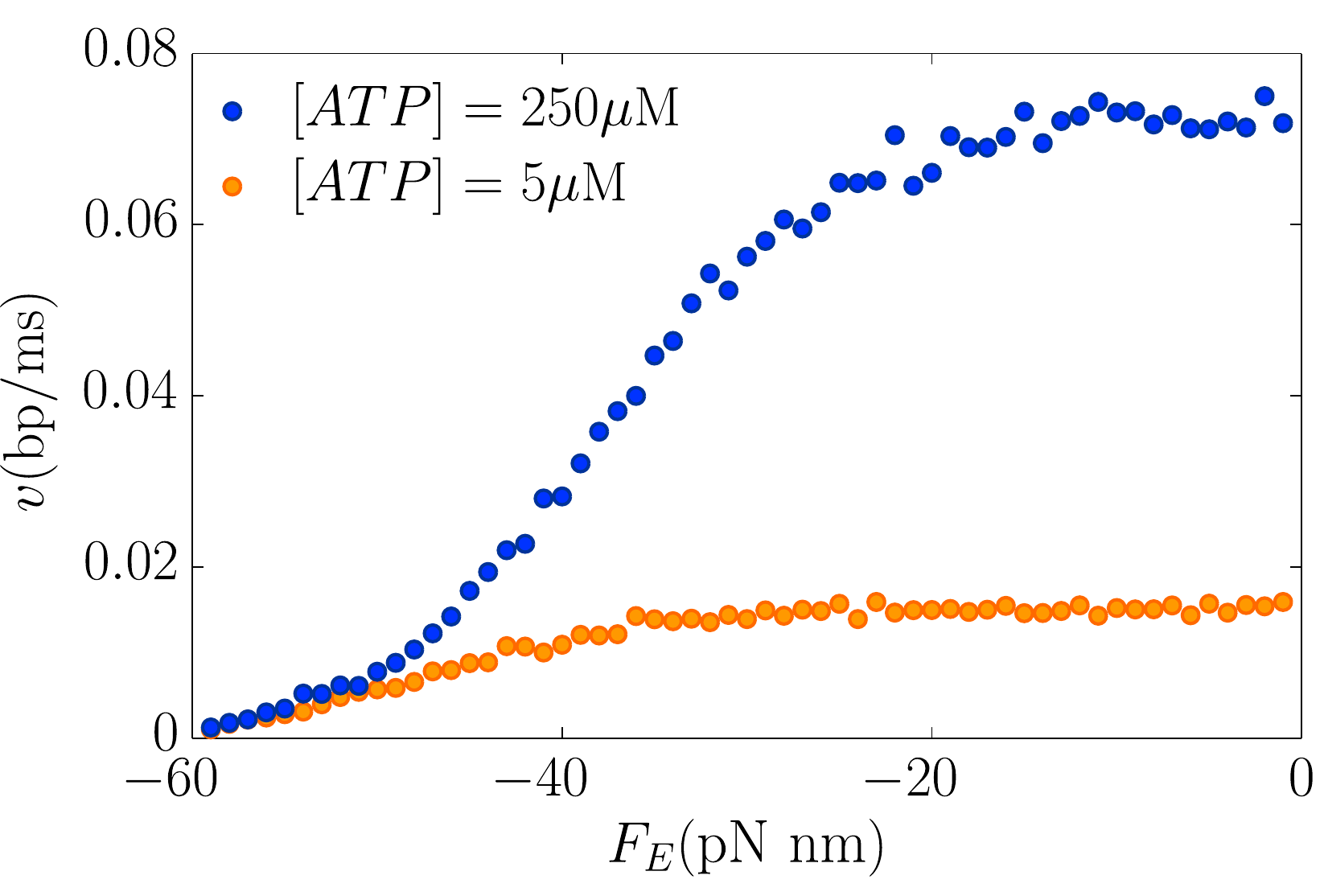}
\caption{\label{velocityfe} (Color Online) Average velocity of the strand versus the external hindering force for different concentrations of ATP.}
\end{figure}

\begin{figure}
\includegraphics[width=\columnwidth]{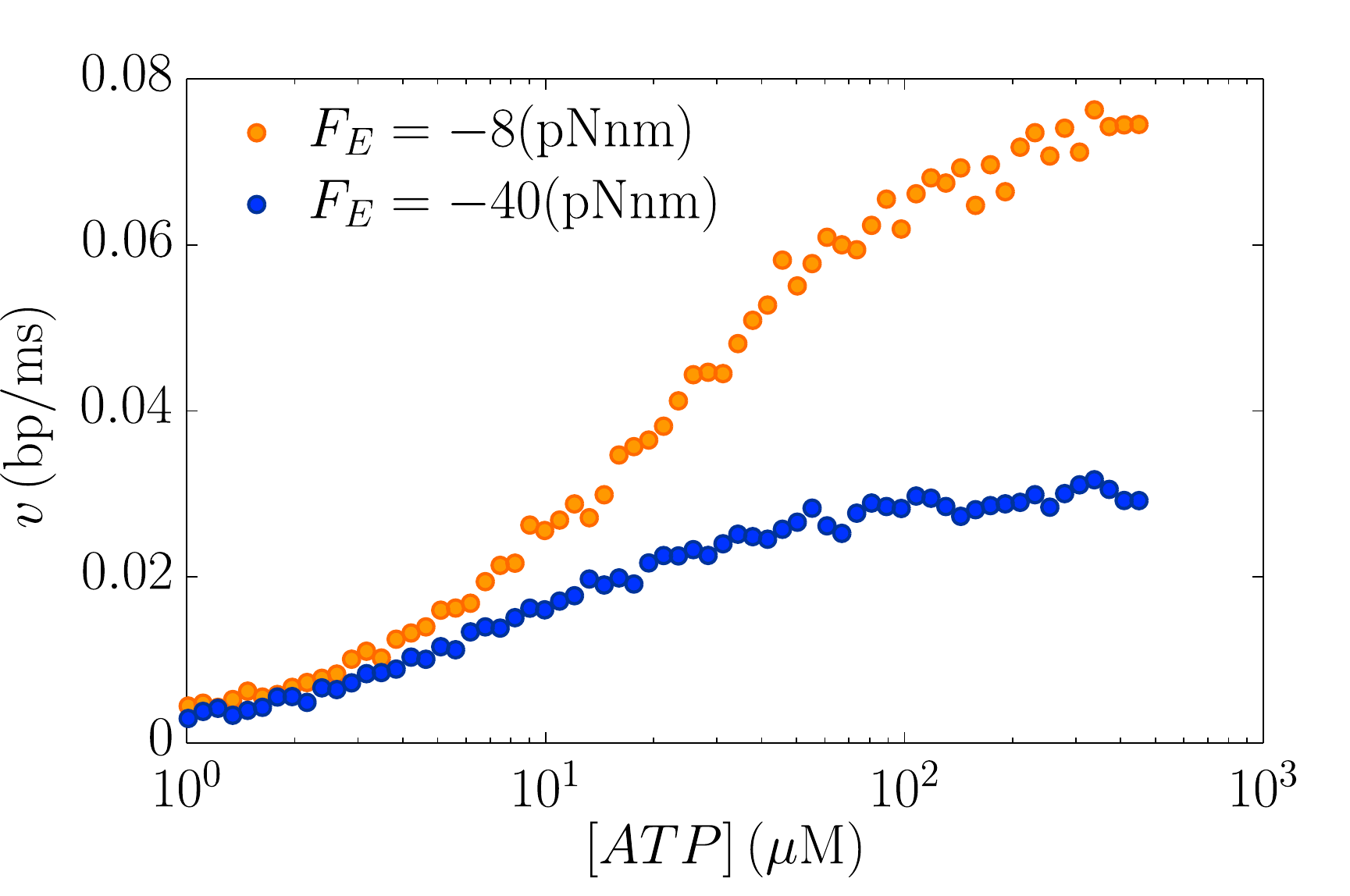}
\caption{\label{velocityatp} (Color Online) Average velocity of the strand versus [ATP] for different values of the external force.}
\end{figure}

From the trajectory, the dependence of the average velocity of the motor $v$ with the external force $F_E$ can be studied obtaining the classical decaying force--velocity curve (Fig.  \ref{velocityfe}). The same occurs when the dependence of the velocity with the ATP concentration is studied (Fig. \ref{velocityatp}) where Michaelis--like curves are seen. The latter evidence has been also found in Ref.~\cite{ajf-damn}, and experimentally reported in Ref.~\cite{Bust05}.

\begin{figure}[h]
\includegraphics[width=\columnwidth]{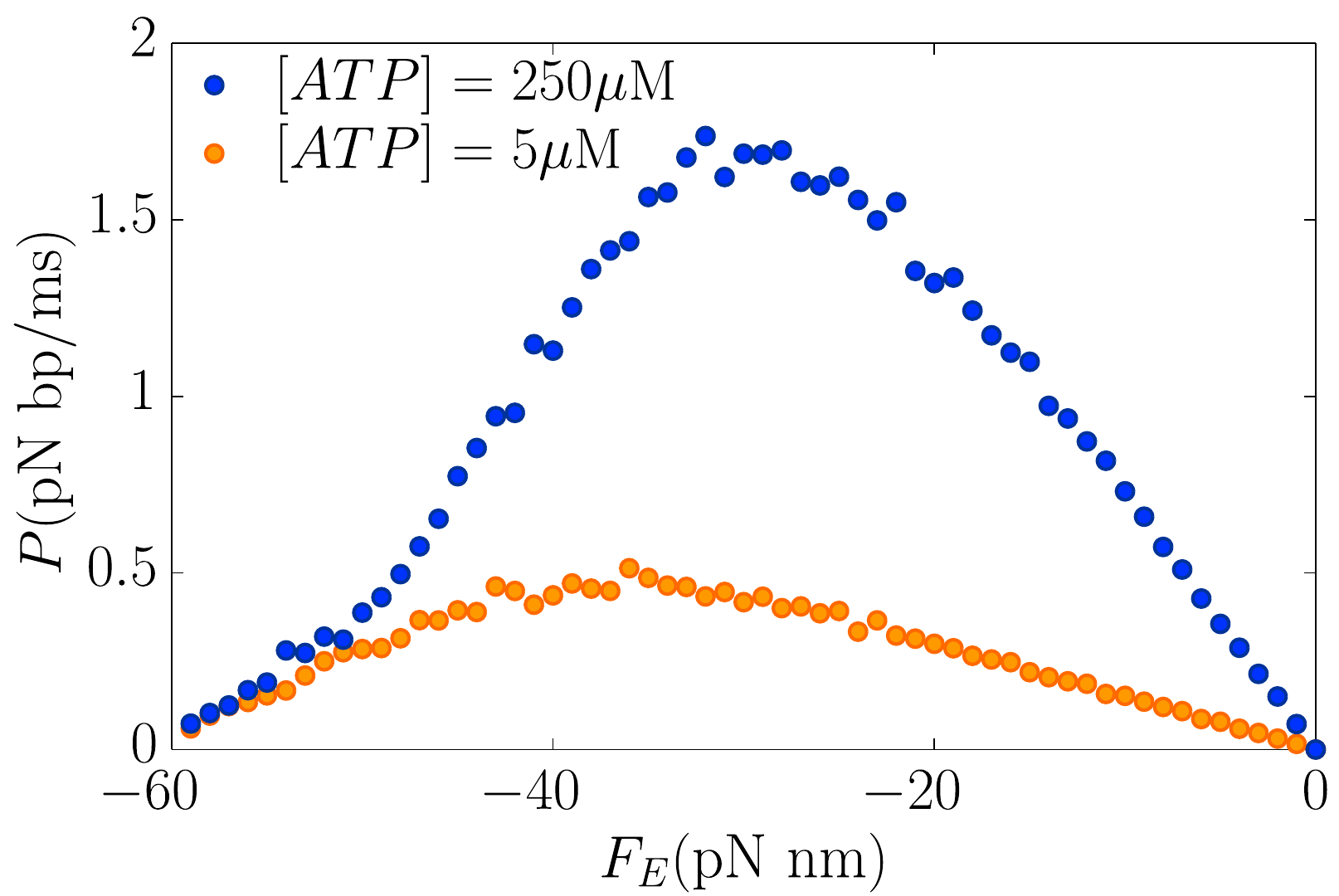}
\caption{\label{figpower} (Color Online) Power of the motor as a function of the external force for different concentrations of [ATP]. }
\end{figure}

The analysis of the dynamics allows also to study the energetics of the model: power and efficiency, where the the power extracted form the motor is defined as $F_E\langle v \rangle$. Fig. \ref{figpower} shows the classical parabolic shape of the power that becomes null in two extreme points, at $F_E=0$, and at the stall force of the motor. This behavior points out an optimum force for which the power extracted from the motor is maximum. It is interesting to note that the curve is not symmetric. Furthermore, the external force needed for a maximum power changes with [ATP].

The efficiency of this motor can also be calculated as the ratio between the output work and the input energy through ATP consumption,
\begin{equation}
 \eta=\frac{F_E\Delta L}{\sum \Delta V_{\mathrm{flash}}}
\end{equation}
being $\Delta L$ the total distance advanced in a trajectory of the motor and $\sum \Delta V_{\mathrm{flash}}$ are all the potential flashing increments corresponding to the same number of hydrolized ATP's of this trajectory \cite{sancho}.

\begin{figure}[b]
\includegraphics[width=\columnwidth]{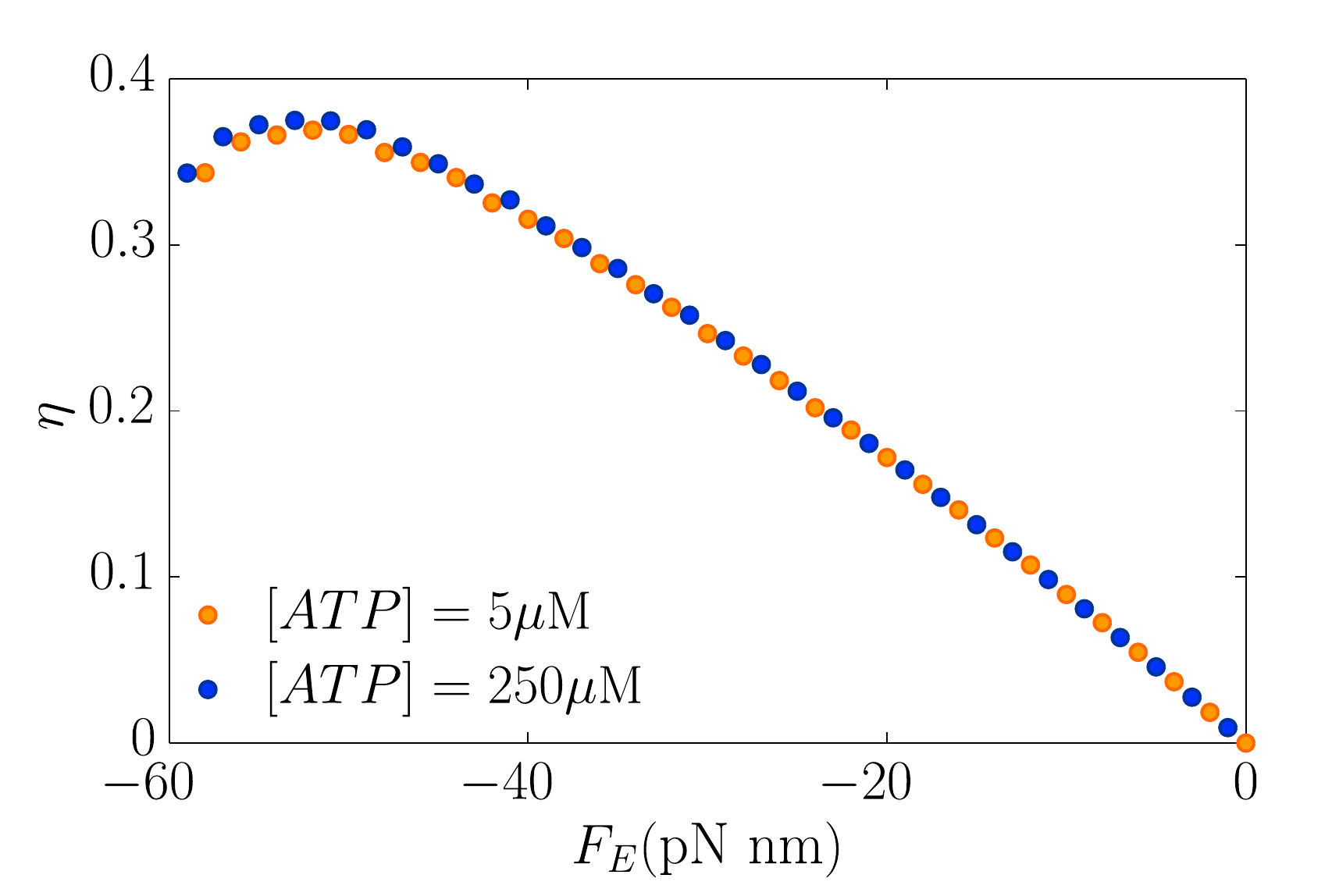}
\caption{\label{figeff} (Color Online) Efficiency of the motor versus the force $F_E$ and two ATP concentrations }
\end{figure}

In Fig. \ref{figeff} one can see the characteristic curve of the efficiency with its maximum at $F_E \sim - 52 {\mathrm{pN}}$, a value close to the stall force. The maximum of efficiency is $\sim 38 \%$, less than most of ideal motors where the efficiency is $50 \%$. It is worth to remark that the efficiency does not depend very much on the ATP concentration and that the maximum of efficiency does not coincide with location of the maximum power.

\section{Comments and conclusions}

In this work we have introduced a model which can explain the trajectory of the $\phi 29$ packaging motor, and in particular its substep features, using an on--off system (flashing potential)  which mimics the ATP hydrolysis processes to produce mechanical work.

The model is based in a careful analysis of the chemical process that occurs in the inner mechanism of the motor, evidenced by the known structure reported in Ref.~\cite{Bust09}. The inner mechanism has been modeled here as four exponentially distributed ATP waiting times, followed by the subsequent ATP hydrolysis events, which give rise to the detailed trajectory sunsteps.

The resulting spatial correlation calculated with different pull forces applied to the chain, shows the typical trend revealed experimentally. Also the velocity of the translocation process for different ATP concentrations, shows a clear agreement with the experimental outcomes, confirming the Michaelis--Menten dependence as a result of the ATP machine functioning.

The authors of \cite{Bust09} report as surprising the fact that the correlation distribution decays with $\Delta x$. In our model this result appears by itself and its origin is due to  missing steps which  induces a lost of the correlation for long distances.

Another interesting point to remark is the presence of high fluctuations observed in the experiments if compared with the ones of the simulations. A possible explanation is that we simulate a rigid system formed by the motor, the DNA chain and the beads with an effective friction constant. Conversely the chain is not rigid, and this may soften the transmission of the motor dynamics to the bead.

As in other investigations \cite{ajf-sin,ajf-rtn,ajf-damn}, we concentrate our analysis on a one--dimensional model. This approach is justified because the polymer translocation in the cited experiments is usually performed with optical traps, and so the polymer is maintained stretched during the process. The polymer is considered rigid, in order to check the features of the machine more than the dissipation given by a smooth chain.

That way we have phenomenologically described the details of the bacteriophage trajectories, using assumption on the motor functioning only. Actually, the model can be extended to other akin motors, such as the recently the study of the packaging motor of bacteriophage $T4$ \cite{kot}. For this motor, pause--unpacking steps in low ATP concentrations and large forces were experimentally observed . Our model can address this issue with the appropriate changes in the biochemical parameters.

\begin{acknowledgments}
This work has been partially supported by the Spanish DGICYT Projects No. FIS2011-25167 and FIS2012-37655-C2-2, co-financed by FEDER funds, and Spanish government fellowship FPU-AP2007-00987(RPC).
\end{acknowledgments}

\end{document}